\documentclass{iaus}
\usepackage{graphicx}

\title[The spectrum of relativistic protons] 
{Study on the  spectrum of the injected relativistic protons}

\author[Y.P. Wang, Y. Lu \& L.Chen]   
{Y.P. Wang $^{1,2}$, %
 Y. Lu $^1$ \and L. Chen$^2$}

\affiliation{$^1$National Astronomical Observatories, Chinese
Academy of Sciences, Beijing 100012,
             China \break email: wangyanping@mail.bnu.edu.cn\\[\affilskip]
$^2$Department of Astronomy, Beijing Normal University, Beijing
100875, China}

\pubyear{2008}
\volume{252}  
\date{?? and in revised form ??}
\jname{Proceedings Title IAU Symposium}
\editors{A.C. Editor, B.D. Editor \& C.E. Editor, eds.}
\begin{document}

\maketitle

\begin{abstract}
About 10\,TeV $\gamma$-ray emission within 10\,pc region from the
Galactic Center had been reported by 4 independent groups.
Considering that this TeV $\gamma$-ray emission is produced via a
hadronic model, and the relativistic protons came from the tidal
disruption of stars by massive black holes, we investigate the
spectral nature of the injected relativistic protons required by the
hadronic model. The calculation was carried on the tidal disruption
of the different types of stars and the different propagation
mechanisms of protons in the interstellar medium. Compared with the
observation data from HESS, we find for the best fitting that the
power-law index of the spectrum of the injected protons is about
-1.9, when a red giant star is tidally disrupted, and the effective
confinement of protons diffusion mechanism is adopted.
\end{abstract}

\firstsection 
\section{Introduction}
The central region of Milky Way is a potential site for the
production of effective particle acceleration and copious
$\gamma$-ray emission. TeV $\gamma$-ray emission from the Galactic
Center(GC) had been reported by 4 independent groups in recent
years: CANGAROO (\cite{Tsuchiya04}), Whipple (\cite{Kosack04}),
HESS(\cite{Aharonian04}), and MAGIC (\cite{Albert06}). One
possibility for this TeV $\gamma$-ray emission source is in the
whole diffuse 10 pc region, which is proposed to be related to the
massive black hole Sgr A$^\ast$ harbored in our Galactic Center
(\cite{Aharonian05}).

There are several radiation mechanisms for the production of TeV
$\gamma$-ray emission. One of these mechanisms proposed that the TeV
$\gamma$-rays can be produced indirectly through the processes of
$\pi$$^0$-decay when relativistic protons are injected into and
interact with the interstellar medium (ISM), called hadronic model
(\cite{Aharonian05}). If this is the case, by assuming that the
initial injected protons' spectrum follows a power-law plus a high
energy cut-off, we can predict the spectrum of the injected
relativistic protons by comparing the spectral energy distribution
(SED) of the TeV $\gamma$-ray emission produced by hadronic model
with observation data from HESS.
\section{Model for the spectrum of the injected relativistic protons}

To calculate the spectrum of the injected relativistic protons, we
consider that the TeV $\gamma$-ray detected by HESS is produced
through the hadronic model, and the injection of the relativistic
protons required by the hadronic model came from the jet of the
black hole Sgr A$^\ast$ when it captures and tidally disrupts a
star(\cite{Lu06}).

The tidal disruption of stars by massive black hole Sgr A$^\ast$
refers to main sequence (MS) stars and red giants. For these two
cases, the total energy carried by the jet is exactly the same as
2.7$\times$10$^{51}$erg, which is believed to be enough for the
production of the injected relativistic protons required by the
hadronic model. The definite difference resulting from the two cases
are the diffusion timescale of the injected protons: 800 yr for the
case of MSs and 2.11$\times$10$^4$yr for the red giants,
respectively (\cite{Lu06}).

Three kinds of diffusion mechanisms are involved for the propagation
of the relativistic protons when they are injected into and interact
with the ISM : (1) the Effective Confinement of Protons (ECP), (2)
the Kolmogorov-Type Turbulence (KTT), (3) the Bohm Diffusion (BD).
For these three mechanisms, the diffusion coefficient depends on the
proton energy, given by the formula of
$D(E)$=10$^{28}$($E$/1$GeV$)$^\delta$$\kappa$ $cm^2s^{-1}$. The case
of ECP corresponds to $\delta$=0.5 and$\kappa$=10$^{-4}$, KTT
corresponds to $\delta$=0.3 and$\kappa$=0.15, and BD corresponds to
$\delta$=1.0 and$\kappa$=10$^{-2}$.

Giving the initial injected relativistic protons spectrum follows a
power-law plus a high energy cut-off,  at a given time of $t$ and
the source distance of $R$, the SED of the TeV $\gamma$-ray emission
produced through the model discussed above can be calculated through
$E_\gamma f(E_\gamma)$=$E_\gamma^2$$V$$q_\gamma(E_\gamma,R,t)$/$S$,
where $E$$_{\gamma}$ is the energy of the emitted $\gamma$-ray
photons, $V$ is the total volume of the ISM, $q$$_{\gamma}$ is the
emissivity of $\gamma$-ray photons, and $S$=4$\pi$($d$$_{ob}$)$^2$,
where $d$$_{ob}$ is the distance from the observation to the source,
and we adopt $d$$_{ob}$=8$kpc$ hereafter.

We address the SED of the emitted $\gamma$-rays in the following
cases: (1) the different power-law index $p$ of injected protons
with values of 1.7, 1.8, 1.9, 2.0, 2.1, 2.2; (2)the MSs and red
giants are tidally disrupted by massive black hole Sgr $A^*$;
(3)three diffusion mechanisms for the cases of ECP, KTT, and BD. By
comparing the theoretical spectrum with the observation data from
HESS within the central 10pc of our Galaxy, we find that the
theoretical energy distribution can fit best with the observation
data when the type of star disrupted by black hole is a red giant
and the diffusion mechanism of protons is the Effective Confinement
of Protons (ECP). In such a case, we derived that the power-law
index of the injected relativistic protons is -1.9.

\section{Conclusions and discussion}

We have investigated the spectrum of the injected relativistic
protons required by the hadronic model to produce the 10 TeV
$\gamma$-ray emission detected by HESS in the center of our Galaxy.
Comparing with the observations, we find that the power-law index of
the initial spectrum of the injected protons should be -1.9. This
result is based on that the tidal disruption of a red giant by the
black hole Sgr $A^*$ is considered and the propagation of protons in
the target gas is the ECP scenario.

\begin{acknowledgments}
This research is supported by the National Natural Science
Foundation of China (Grants 10573021 and 10778716).
\end{acknowledgments}

\end{document}